\documentclass[review,numbers,sort,compress]{elsarticle}

\usepackage{lineno}
\usepackage{graphicx}
\usepackage{latexsym}%
\usepackage[table]{xcolor}
\usepackage{amsfonts}
\usepackage{booktabs}
\usepackage{lmodern}
\usepackage{lineno}
\usepackage{soul}
\usepackage{enumitem}
\usepackage{comment}
\usepackage{relsize}
\usepackage{lipsum}
\usepackage{color}
\usepackage{siunitx}
\usepackage[cal=boondoxo]{mathalfa}
\usepackage[breaklinks=true]{hyperref}
\usepackage{breakcites}
\usepackage{blindtext, rotating}
\usepackage{framed} 
\usepackage{multicol} 
\usepackage{pifont}
\usepackage[parfill]{parskip}
\usepackage{nomencl}
\usepackage{colortbl}
\usepackage{tabularx}
\usepackage[defaultlines=3,all]{nowidow}
\usepackage{amssymb}
\usepackage{amsmath}
\usepackage{epstopdf}
\usepackage{setspace}
\usepackage{xspace}
\usepackage{subcaption}
\usepackage{epsfig}
\usepackage{placeins}
\usepackage{epstopdf}
\usepackage{wrapfig}
\usepackage{algorithm}
\usepackage[noend]{algpseudocode}
\usepackage{caption}
\usepackage[version=3]{mhchem}
\usepackage{lineno,hyperref}
\usepackage{cleveref}
\usepackage[font=small,skip=0pt]{caption}
\usepackage{geometry}
\usepackage{todonotes}
\usepackage[normalem]{ulem} 

\modulolinenumbers[5]
\definecolor{lavender}{rgb}{0.75, 0.58, 0.89}

\newcolumntype{M}[1]{>{\centering\arraybackslash}p{#1}}
\newcolumntype{P}[1]{>{\raggedright\arraybackslash}p{#1}}

\setlist[itemize]{leftmargin=*}
\newif\ifcomments 
\commentsfalse
\commentstrue 

\definecolor{Green}{rgb}{0,0.5,0}

\definecolor{lightgray}{gray}{0.9}
\definecolor{amethyst}{rgb}{0.8, 0.0, 0.8}
\definecolor{aogreen}{rgb}{0.01, 0.75, 0.24}

\journal{Journal of \LaTeX\ Templates}

\begin{document}
\begin{frontmatter}
\title{Skeletal Reaction Models for Gasoline Surrogate Combustion}


\author[mymainaddress]{Yinmin Liu}
\author[mymainaddress]{Hessam Babaee}
\author[mymainaddress]{Peyman Givi}
\author[mysecondaddress]{Daniel Livescu}
\author[myteritaryaddress]{Arash G.\ Nouri \corref{mycorrespondingauthor}}

\address[mymainaddress]{Department of Mechanical Engineering and Materials Science, University of Pittsburgh, Pittsburgh, PA 15261, USA}
\address[mysecondaddress]{Los Alamos National Laboratory, Los Alamos, NM 87544, USA}
\address[myteritaryaddress]{Aurora Corporation, Mountain View, CA 94043, USA}

\cortext[mycorrespondingauthor]{Corresponding author \\ E-mail address: arash.nouri@pitt.edu.}

\begin{abstract}
Skeletal reaction models are derived for a four-component 
gasoline surrogate model via an instantaneous local sensitivity analysis technique.
The sensitivities of the species mass fractions and the temperature with respect to the reaction rates are estimated by a reduced-order modeling (ROM) methodology. 
Termed ``implicit time-dependent basis CUR (implicit TDB-CUR),'' this methodology is based on the CUR matrix decomposition and incorporates implicit time integration for evolving the bases.
The estimated sensitivities are subsequently analyzed to develop skeletal reaction models with a fully automated procedure. The 1389-species gasoline surrogate model developed at Lawrence Livermore National Laboratory (LLNL) is selected as the detailed kinetics model. The skeletal reduction procedure is applied to this model in a zero-dimensional constant-pressure reactor over a wide range of initial conditions.
The performances of the resulting skeletal models are appraised by comparison against the results via the LLNL detailed model, and also predictions via other skeletal models.
Two new skeletal models are developed consisting of 679 and 494 species, respectively. The first is an alternative to an existing model with the same number of species. The predictions with this model reproduces the detailed models vital flame results with less than 1\% errors. The errors via the second model are less than 10\%.

\end{abstract}
\end{frontmatter}

\section{Introduction}
\label{section:Intro} 

Gasoline fuels consist of hundreds of hydrocarbons. Computational modeling and simulation of such complex chemical mixtures require accounting for numerous species 
mass fractions. To simplify this, gasoline surrogate mixtures are formulated with fewer hydrocarbons. These surrogates mimic the chemical kinetics and thermodynamic properties of real gasolines \cite{zhen17ove, sarathy18rec}. However, surrogate models still involve $\mathcal{O}$(1000) species, creating significant computational challenges, especially for multidimensional reacting flow simulations. Therefore, developing skeletal kinetics models for gasoline surrogates by eliminating unimportant species and reactions from the detailed models is essential for feasible simulations.

During the last two decades, a variety of techniques have been developed for skeletal kinetics reduction, by systematical eliminations of unimportant species and reactions while maintaining the overall predictive capability~\cite{law10com, de16rev, GM11}. Local sensitivity analysis (LSA), reaction flux analysis~\cite{SCGJ10,wang16imp}, computational singular perturbation (CSP) analysis \cite{lam89und,lam93usi,lam94csp,lu01com}, and directed relation graph (DRG) and its variants~\cite{LL05,ZLL07,LL09,NSR10} have often been utilized. 
The LSA explores the response of model outputs to small changes in parameters from their nominal values~\cite{Shuang16}. Global sensitivity analysis considers uncertainty in the kinetic parameters (\textit{i.e.}collision frequencies and activation energies) and non-linear coupling effects~\cite{Turanyi90b,Sobol90,HS96,RA99,Sobol01,LSR02,Saltelli04,SYWL09,ESC12}. 
Instantaneous sensitivity analysis evaluates sensitivities at a specific time in a dynamic system, while time-independent sensitivity analysis measures how values change with parameter variations in a time-averaged sense~\cite{lakrisenko24ben, lakrisenko23eff}. Instantaneous LSA-based techniques, which are computationally costly for large kinetics models, include methods such as the principal component analysis (PCA) ~\cite{BLK97,EC11,PSDTS11,PS13,ME14,CIGP16,MICSP18}, and species ranking construction~\cite{SFCFR16}. 
The f-OTD method is specifically designed for low-rank approximations in instantaneous local sensitivity calculations. In Refs.~\cite{NBGCL22,LBGCLN24,nouri24ske} the f-OTD is employed to approximate sensitivities to generate skeletal models for hydrocarbon combustion and nuclear burning. In order to enhance the computational advantages of f-OTD and to overcome its numerical challenges, a new method termed ``implicit time-dependent basis CUR (implicit TDB-CUR)'' has been developed~\cite{DPNFB23,naderi25cro}. The CUR denotes the CUR matrix decomposition, which constructs a low-rank approximation matrix in terms of a subset of its columns (C), a core matrix (U), and a subset of its rows (R)~\cite{goreinov1997the,mahoney2009cur}. The implicit TDB-CUR method combines sparse sampling, CUR decomposition, and a cost-efficient implicit time integration algorithm, offering improved efficiency and stability compared to f-OTD. These enhancements are particularly useful for dealing with stiff reaction networks characterized by reactions with orders of magnitude different time scales. Gasoline surrogate models are among these reaction networks. 

The Lawrence Livermore National Laboratory (LLNL) gasoline surrogate model \cite{mehl11det} contains the chemistry parameters for various hydrocarbon compounds including iso-octane, n-heptane, toluene, and C5-C6 olefins. This model has been validated against a wide range of experimental data, making it broadly applicable for predicting various gasoline surrogate combustion phenomena~\cite{mehl11det,kuk2012exp,kuk2013aut}. However, with 1389 species and 10734 irreversible reactions, the model is computationally prohibitive for practical simulations. 
Mehl et al. \cite{mehl2011red} developed a 312-species model using a computer-assisted reduction mechanism, followed by a search algorithm that sequentially assessed the importance of each species. Despite its compact size, the model is not sufficiently accurate (see Section \ref{subsection:model_validation}). Lu \textit{et al.}~\cite{lu09tow} constructed a reasonably accurate 679-species skeletal model but still left room for further size reduction. Niemeyer \textit{et al.}~\cite{niemeyer15red} derived two reduced models containing 97 and 79 species, each tailored to different operating conditions. Their approach incorporated a directed relation graph with error propagation and sensitivity analysis (DRGEPSA)~\cite{NSR10}, unimportant reaction elimination, isomer lumping, and analytical quasi-steady-state (QSS) assumption. While isomer lumping and QSS significantly reduce model size, they compromise model versatility by neglecting isomer-specific reactivity differences and assuming rapid equilibration of certain species.

The objective of this work is to develop accurate skeletal models from the LLNL gasoline surrogate model using sensitivities estimated via the implicit TDB-CUR method. 
This reduction targets a four-component gasoline surrogate (iso-octane, n-heptane, 2-pentene, and toluene), formulated by Mehl \textit{et al.}~\cite{mehl2011red} and focuses on a lean-to-stoichiometric, low-temperature, homogeneous charge compression ignition (HCCI)-like range of conditions. The composition of the surrogate fuel is given in Table~\ref{tbl:composition}. 
The models are evaluated alongside two previously developed reduced models, and the detailed model to predict the ignition delay, the flame speed, and the flame structure. The efficiency of the implicit TDB-CUR sensitivity approximation is also assessed via comparison against the full-size sensitivity integration. 

\newcolumntype{?}{!{\vrule width 1.0pt}}
\begin{table}
\caption{Composition of the Surrogate Fuel}
\label{tbl:composition}
\scriptsize
\centering
\begin{tabular}{?M{2.4cm}?M{2.4cm}?}
\specialrule{1.0pt}{0pt}{0pt}
Components &  Mole Fractions \\ 
\specialrule{1.0pt}{0pt}{0pt}
iso-octane   & 48.70\%   \\ \hline
n-heptane & 15.45\%  \\ \hline
toluene  & 30.60\%  \\ \hline
2-pentene & 5.25\% \\ \hline
\specialrule{1.0pt}{0pt}{0pt}
\end{tabular}
\end{table}


\section{Methodology}\label{section:method}
The reduction procedure consists of the followings: First, the implicit TDB-CUR method with the fourth-order backward differentiation formula (BDF4)~\cite{curtiss52int,gear1967num,suli2003int} is used to compute sensitivities across multiple initial conditions (cases). These sensitivities are then collected and analyzed to produce a ranking of the species and the reactions. Finally, reduced models are constructed based on these rankings.

\subsection{Sensitivity Analysis}

As described in Refs.~\cite{ZZT03,NBGCL22,LBGCLN24}, local linear sensitivity coefficients in a chemical kinetics system is evaluated by infinitesimal perturbations of all reaction rates. The full-size matrix containing all the sensitivity coefficients evolves based on the following matrix differential equation:
\begin{equation}
 \label{eq:sensi}
 \begin{split}
\frac{dS(t)}{dt} = L(t)S(t) + F(t),
\end{split}
\end{equation}
where $S(t) \in \mathbb{R}^{n_{eq} \times n_{rc}}$ is the sensitivity matrix, $L(t) \in \mathbb{R}^{n_{eq} \times n_{eq}}$ is the Jacobian matrix, and $F(t) \in \mathbb{R}^{n_{eq} \times n_{rc}}$ is the forcing term. The number of reactions is noted as $n_{rc}$. The number of state variables is $n_{eq} = n_{sp} + 1$ where $n_{sp}$ denotes the number of chemical species. Temperature is also solved as a state variable. Equation (\ref{eq:sensi}) represents the full-order model (FOM).
The implicit TDB-CUR method aims to accurately approximate the sensitivity matrix $S^k = S(t^k)$ at all time steps. The low-rank components are stored in a singular value decomposition (SVD)-like factorized form: $S^{k} = U^{k} \Sigma ^{k} Y^{k}$. To compute the low-rank components for the next time step $t^{k+1}$, the CUR algorithm is applied, requiring the solution of selected columns and rows of the sensitivity matrix.

In CUR decomposition, the sensitivity matrix $S(t) \in \mathbb{R}^{n_{eq} \times n_{rc}}$ is approximated by a subset of its columns and rows, $S \approx C U R$,  where $C \in \mathbb{R}^{n_{eq} \times r}$ consists of $r$ selected columns of $S$, $R \in \mathbb{R}^{r \times n_{rc}}$ consists of $r$ selected rows of $S$, and $U \in \mathbb{R}^{r \times r}$ is a low-rank coefficient matrix that captures the interaction between the selected columns and rows. Unlike singular value decompositions, CUR provides a more interpretable decomposition, as the selected columns and rows correspond directly to actual sensitivity information rather than abstract linear combinations. In the implicit TDB-CUR, this decomposition is applied to the sensitivity matrix at each time step, maintaining a low-rank representation without requiring additional evolving equations for the low-rank components. The columns and rows are integrated implicitly using the BDF4 method. The updated sensitivities are stored in the same format: $S^{k+1} = U^{k+1} \Sigma ^{k+1} Y^{k+1}$.
To obtain the initial conditions of the low-rank components, the FOM is integrated for several time steps. Then the full-order sensitivity matrix $S^o = S(t^o)$ is factorized by a SVD. This factorization leads to three low-rank components: $U^o \in \mathbb{R}^{n_{eq} \times r}$, $\Sigma ^o \in \mathbb{R}^{r \times r}$, and $Y^o \in \mathbb{R}^{n_{rc} \times r}$, such that $S^o \approx U^o\Sigma^o{Y^o}^T$, where $r$ denotes the number of retained modes ($r \ll n_{eq}$). Then, instead of integrating the FOM, the three low-rank components are updated and saved at each time step. In this way, the computing and memory cost of implicit sensitivities integration is substantially reduced in comparison to that required for the solution of the FOM.

The integration of selected columns and rows involves solving the following equations:
\begin{subequations}
 \label{eq:update}
 \begin{eqnarray}
  S^{k+1}(:,I_c) = S^{k}(:,I_c) + \delta S^k(:,I_c), \\
  S^{k+1}(I_r,:) = S^{k}(I_r,:) + \delta S^k(I_r,:),
\end{eqnarray}
\end{subequations}
where $\delta S^k(:,I_c)$ and $\delta S^k(I_r,:)$ denote the correction terms for the columns $I_c$ and rows $I_r$, respectively. Matlab-style indexing is adopted, where $S^k(:,I_c)$ denotes a submatrix containing the specified columns, and $S^k(I_r,:)$ denotes a submatrix containing the specified rows. The GappyPOD+E method \cite{PDG20} is used to select $I_c$ and $I_r$ to minimize residuals from the low-rank approximation.
The column correction term is obtained by solving the linear system:
\begin{equation}
 \label{eq:linear1}
 \begin{split}
  A^{k} \: \delta S^k(:,I_c) = b^k,
\end{split}
\end{equation}
where $A^k$ and $b^k$ depend on the implicit integrator. For the BDF4 scheme: $A^k = (I - \frac{12}{25}\Delta tL)$ and $b^k = [\frac{48}{25}S^k(:,I_c) - \frac{36}{25}S^{k-1}(:,I_c) + \frac{16}{25}S^{k-2}(:,I_c) - \frac{3}{25}S^{k-3}(:,I_c)] - S^{k+1}_i(:,I_c) + \frac{12}{25}\Delta t(L^kS^k(:,I_c)+F^k(:,I_c))$, where $I$ denotes the identity matrix. Since the local sensitivity system is linear, only one linear solve is required. For the FOM, the same formula is used, except that $I_c$ includes all of the columns.

Row integration requires a special treatment. In a linear system of equations, the solution for any entry in the unknown vector is coupled to all other entries in that vector. Therefore, integrating the selected rows requires solving the entire linear system. To reduce the computational cost, a low-rank approximation of $\delta S^k(I_r,:)$ is employed in implicit TDB-CUR. Resulting a reduced-order linear system:
\begin{equation}
 \label{eq:linear2}
 \begin{split}
  A^k_{(r)} \: \delta S^k(I_r,:) = b^k_{(r)},
 \end{split}
\end{equation}
where $A^k_{(r)} = A^k(I_r,I_r) + A^k(I_r,\tilde{I_r})U^k_{\sigma}(\tilde{I_r},:)U^k_{\sigma}(I_r,:)^\dag$ and $b^k_{(r)} = [\frac{48}{25}S^k(I_r,:) - \frac{36}{25}S^{k-1}(I_r,:) + \frac{16}{25}S^{k-2}(I_r,:) - \frac{3}{25}S^{k-3}(I_r,:)] - S^{k+1}_i(I_r,:) + \frac{12}{25}\Delta t(L^k(I_r,:)S^k+F^k(I_r,:))$. $\tilde{I_r}$ denotes the complement set of $I_r$ and the $\dag$ sign denotes a pseudo-inverse. The orthonormal basis $U^k_{\sigma}$ is obtained by a SVD operation on a matrix $S^k_{\sigma}$ that is obtained by subtracting previous column solutions from the $t^{k+1}$ column solution: $S^k_{\sigma} = S^{k+1}(:I_c) - S^{k}(:I_c) - S^{k-1}(:I_c) - S^{k-2}(:I_c) - S^{k-3}(:I_c)$. Equation (\ref{eq:linear2}), with $A^k_{(r)} \in \mathbb{R}^{r \times r}$, can be solved efficiently to find $\delta S^{k}(I_r,:)$. 

The low-rank components of sensitivity matrix at $t^{k+1}$ is computed as follows:
\begin{subequations} \label{eq:cur}
\begin{eqnarray}
S^{k+1}(:,I_c) &=& U\Sigma Y^T, \label{eq:cur0} \\
V &=& [U(I_r,:)^\dag \: S^{k+1}(I_r,:)]^T, \label{eq:cur1} \\
V &=& Y^{k+1} \Sigma^{k+1} R_U,  \label{eq:cur2} \\
U^{k+1} &=& UR_U,  \label{eq:cur3}
\end{eqnarray}
\end{subequations}
\noindent where $U$ is computed via a SVD on column solutions. $\Sigma^{k+1}$ and $Y^{k+1}$ are obtained by a SVD on the matrix $V$. Finally, the remaining sensitivity component $U^{k+1}$ is obtained by rotating the basis $U$.

\subsection{Reactions and Species Ranking}
After collecting low-rank sensitivity components from multiple cases, the reactions and the species can be ranked. The right singular vectors $Y^k = Y(t^k)$, which reflect the reaction importance, is weighted by the singular values $\Sigma^k$:
\begin{equation}
 \label{eq:concatenate1}
 \begin{split}
  W^{k}_{(j)} = \sum_{n} |Y^k_n| \Sigma ^k_n / \sum_{n} \Sigma ^k _n  ,
 \end{split}
\end{equation}
\noindent where the index $n$ denotes the n$th$ right singular vector or singular value. Each element of $W^{k}_{(j)} \in \mathbb{R}^{n_{rc}} $, is positive and is associated with a certain reaction. The larger the associated $W$ value, the more important the reaction is. The parameter $\chi \in \mathbb{R}^{n_{rc}}$ is defined as the highest value associated with $W$ across all resolved time-steps and cases. The elements of the $\chi$ vector are sorted in descending order to find the indices of the most important reactions in the detailed model. The species are ranked based on their first presence in the sorted reactions, \textit{i.e.} a species that first appears in a higher-ranked reaction is considered more important than one that first participates in a lower-ranked reaction. This process results in a reaction and species ranking based on the $\chi$ vector.
The reduced model is constructed by removing the less important species from the species ranking while retaining all reactions associated with the remaining species. 

The skeletal reduction scheme as shown here is based on instantaneous local sensitivities. Unlike PCA-based sensitivity methods~\cite{EC11}, where full sensitivities are computed first and then reduced, the implicit TDB-CUR approach solves for the sensitivities directly in their reduced form. This leads to computational savings in floating-point operations (FLOPs) and substantial reductions in memory usage, while also eliminating the need to store full-size sensitivity data.

\section{Skeletal Reduction of Gasoline Surrogate}
The implicit TDB-CUR method with the BDF4 integrator~\cite{suli2003int} is employed for low-rank approximation of gasoline surrogate model sensitivities. A total of 11 cases are considered, with initial conditions summarized in Table \ref{tbl:cases} taken from Refs.~\cite{mehl2011red,niemeyer15red}. These cases cover the ranges of $750$ $K$ to $1200$ $K$ for the temperature, $10$ $atm$ to $60$ $atm$ for the pressure, and $0.2$ to $1.0$ for the equivalence ratio. The CUR rank is set as $r = 7$. 
To benchmark the efficiency and the accuracy of implicit TDB-CUR, the FOM is also solved for all the cases by the BDF4. Following the FOM solution, an SVD of the sensitivity matrix is performed at each time step. The data from these SVD is referred to as the instantaneous SVD (iSVD). The seven largest singular values from the iSVD are compared with those obtained using the implicit TDB-CUR. The sensitivities estimated by the implicit TDB-CUR are further analyzed to generate species and reaction rankings. The performance of the reduced models is evaluated by comparison against the results of the detailed model. Two existing reduced models are also included for comparison: $chem312$ \cite{mehl2011red} and $chem679$ \cite{lu09tow}.

\begin{table}
\caption{Clock Times for Computing Sensitivities}
\label{tbl:cases}
\scriptsize
\centering
\begin{tabular}{?M{1.2cm}?M{2.4cm}?M{1.4cm}?M{2.0cm}?M{2.4cm}?}
\specialrule{1.0pt}{0pt}{0pt}
& Initial Conditions & Time Steps & FOM & Implicit TDB-CUR\\ 
\specialrule{1.0pt}{0pt}{0pt}
Case 1 & T800, P10, $\phi$1.0 & 6466 & 6320.6 sec & 3587.3 sec\\ \hline
Case 2 & T750, P60, $\phi$1.0 & 6826 & 6905.2 sec & 3874.6 sec\\ \hline
Case 3 & T1200, P60, $\phi$0.6 & 4544 & 3633.8 sec & 2080.0 sec \\ \hline
Case 4 & T1100, P10, $\phi$0.6 & 4766 & 3891.5 sec & 2210.6 sec\\ \hline
Case 5 & T1000, P60, $\phi$0.6 & 4948 & 4538.6 sec & 2575.6 sec\\ \hline
Case 6 & T800, P10, $\phi$0.6 & 6566 & 6239.1 sec & 3593.2 sec\\ \hline
Case 7 & T750, P10, $\phi$0.6 & 6820 & 6629.3 sec & 3842.9 sec\\ \hline
Case 8 & T750, P60, $\phi$0.6 & 6764 & 6774.1 sec & 3847.3 sec\\ \hline
Case 9 & T800, P60, $\phi$0.2 & 4440 & 3031.3 sec & 1745.5 sec\\ \hline
Case 10 & T700, P20, $\phi$0.2 & 4852 & 3209.0 sec & 1838.3 sec\\ \hline
Case 11 & T800, P20, $\phi$0.2 & 4244 & 3435.2 sec & 1956.3 sec\\ \hline
\specialrule{1.0pt}{0pt}{0pt}
\end{tabular}
\end{table}

\subsection{TDB-CUR Evaluation}
\begin{figure}[!ht]
\centering 
 \includegraphics[width=8cm]{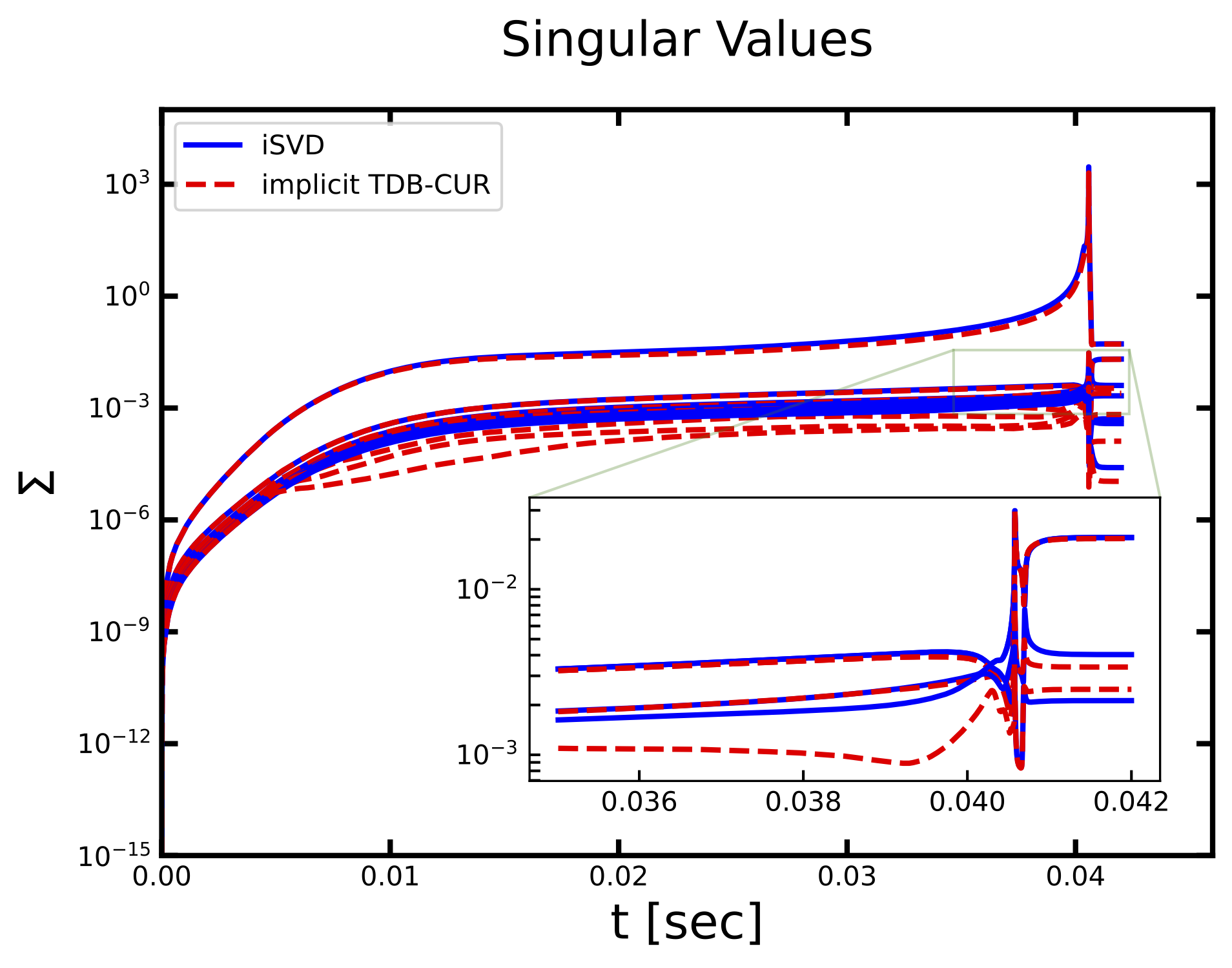}
 \caption{Singular values in case 1 ($\phi$ = 1.0, T=800K, P=10 atm).}
\label{FIG:singular_values}
\end{figure}

Figure \ref{FIG:singular_values} provides a comparison of the singular values of the sensitivity matrix obtained by implicit TDB-CUR against those via iSVD for Case 1. It is shown that there is one dominant mode for which the singular value is multiple orders of magnitude larger than the others. The implicit TDB-CUR is able to accurately capture the first three singular values. The smaller modes do not align with the FOM results, but have no significant effects on the predicted results, since the right singular values are weighted by the singular values. As for the efficiency of the implicit TDB-CUR, the clock time of ROM simulation is compared against FOM in Table \ref{tbl:cases}. In all the cases, the ROM requires less than 60\% time of the FOM. 

\subsection{Skeletal Model Validation}
\label{subsection:model_validation}
\begin{figure}[!ht]
\centering 
 \includegraphics[width=16cm]{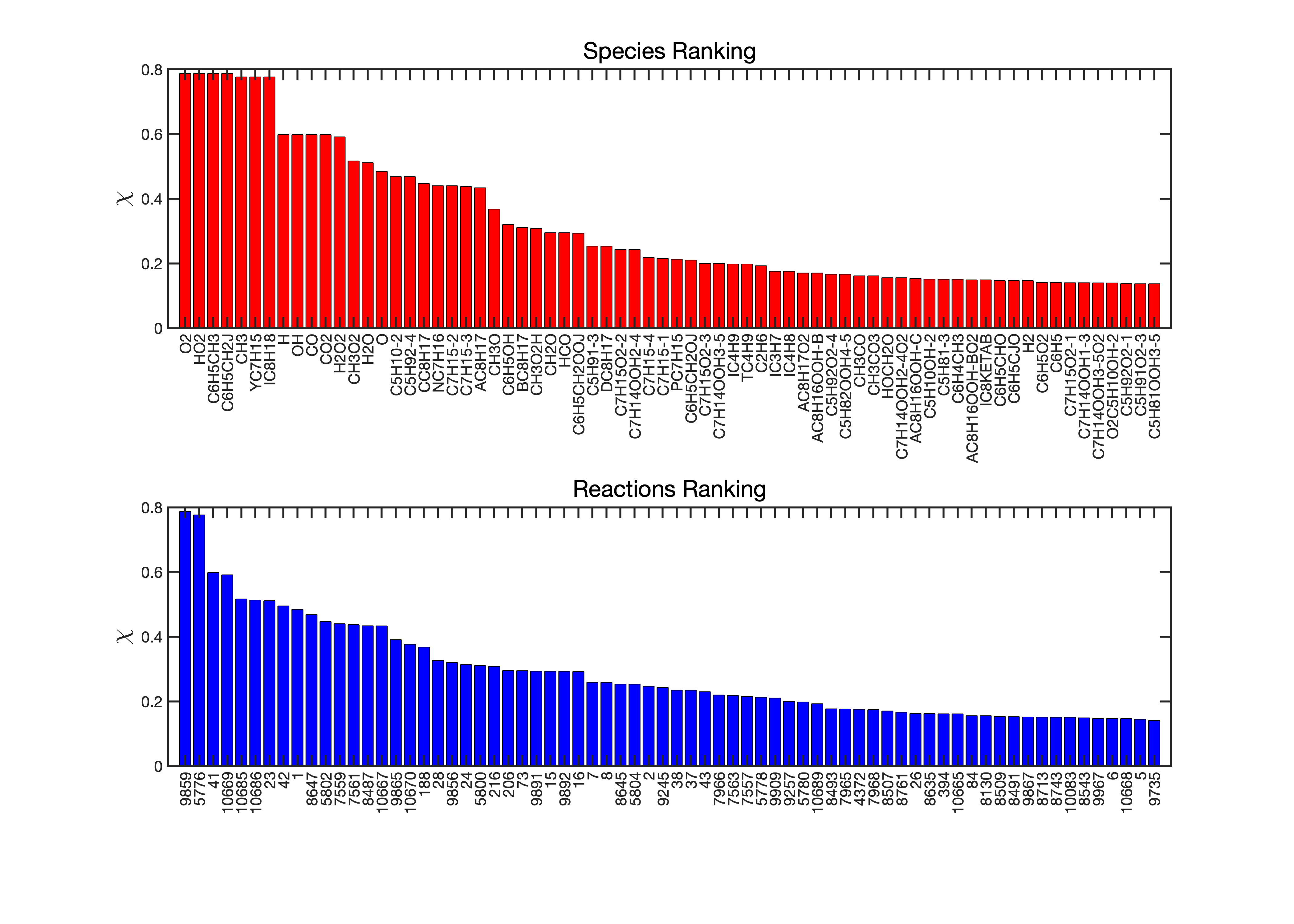}
 \caption{Importance ranking of the species and reactions. Reaction ranking is labeled by the reaction indices associated with the supplemental Cantera file}
\label{FIG:ranking}
\end{figure}

In Fig.~\ref{FIG:ranking}, the top $70$ most important species and reactions of the LLNL gasoline surrogate model are ranked based on their importance. The higher the $\chi$ values, the higher the importance of a species/reaction. The Appendix provide the complete ranking of the species and rankings. 
In the reactions ranking, a toluene oxidization reaction ($Reaction$ $9859:$ $C_6H_5CH_3$ + $O_2$ $\rightarrow$ $C_6H_5CH_2J$ + $HO_2$) is identified as the most important. The species involved in this reaction are identified as the most important species, as indicated by the species ranking. The second most important is an iso-octane pyrolysis reaction ($Reaction$ $5776:$ $IC_8H_{18}$ $\rightarrow$ $CH_3$ + $YC_7H_{15}$). This is consistent with the findings in Ref.~\cite{ghiasi2019gas}.
Based on the species ranking two reduced reaction models are suggested. The first contains the top 679 most important species ($rd679$), the same number as that in the $chem679$~\cite{lu09tow}. This model yields less than \%1 error. The second model with the top 494 species ($rd494$) guarantees 10\% maximum relative error of all comparative assessments with the minimum number of species. 
The numbers of reactions for these reduced models are listed in Table \ref{tbl:mechanisms}. The reaction size refers to the number of irreversible reactions.

\begin{table}
\caption{Chemistry Models}
\label{tbl:mechanisms}
\scriptsize
\centering
\begin{tabular}{?M{2.4cm}?M{2.4cm}?M{2.4cm}?}
\specialrule{1.0pt}{0pt}{0pt}
Model & $n_{sp}$ & $n_{rc}$\\ 
\specialrule{1.0pt}{0pt}{0pt}
Detailed   & 1389 & 10734\\ \hline
rd679   & 679 & 5731 \\ \hline 
rd494   & 494 & 4154 \\ \hline 
\specialrule{1.0pt}{0pt}{0pt}
\end{tabular}
\end{table}

\begin{figure}[h!]
\centering 
 \includegraphics[width=16cm]{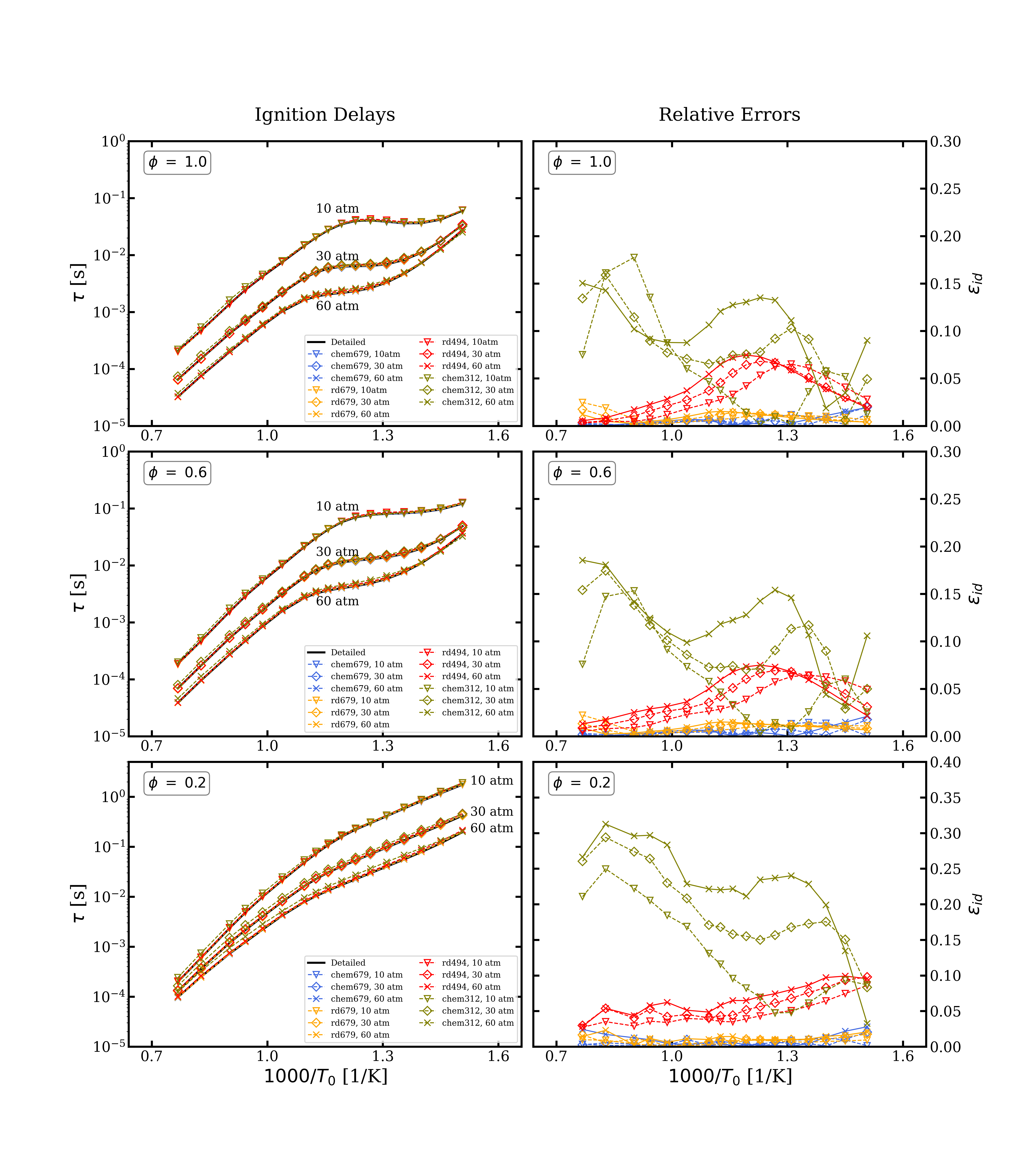}
 \caption{Left: Ignition delay test of the reduced models. Right: Relative errors $\epsilon_{id} = |\tau_{reduced}-\tau_{detailed}| / \tau_{detailed}$.}
\label{FIG:ignition}
\end{figure}

In Fig.~\ref{FIG:ignition}, the ignition delays are shown for several equivalence ratios and pressure values. The relative errors in predicting the delays are also provided and indicate that $rd679$ and $chem679$ provide the best overall accuracy. The relative errors associated with these two models are of order 1\%. For the $rd494$ model, the relative errors are less than 8\% in most cases. 
With $\phi$ $=$ $0.6$ $\&$ $1.0$, the $rd494$ models show larger errors when $1000/T_0$ $=$ $1.1 \sim 1.4$ $K^{-1}$, which correspond to regions with lower gradients. 
This behavior is known as the ``negative temperature coefficients'' (NTC) \cite{law2012ntc}. The shape of the $rd494$ relative error curves indicates that the NTC may increase the difficulty of producing accurate reduced models using sensitivity coefficients.
The chem312 model \cite{mehl2011red} provides errors up to 30\%, but yields accurate predictions when $P$ $=$ $10$ $atm$ and NTC behavior occurs. 

\begin{figure}[!ht]
\centering 
 \includegraphics[width=16cm]{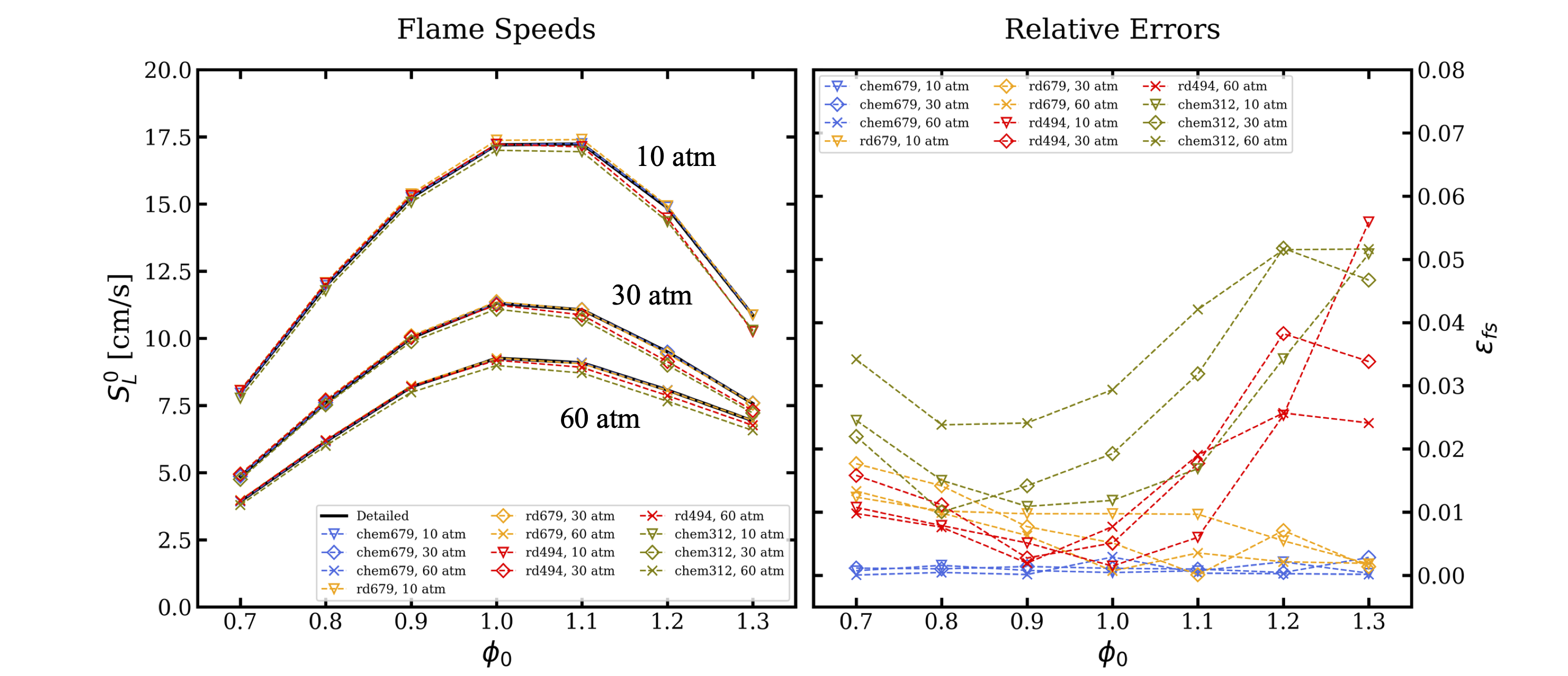}
 \caption{Left: Laminar flame speed test ($T_0 = 300$ $k$). Right: Relative errors $\epsilon_{fs} = |S^0_{L,reduced}-S^0_{L,detailed}| / S^0_{L,detailed}$.}
\label{FIG:flamespeed}
\end{figure}

\begin{figure}[!ht]
\centering 
 \includegraphics[width=16cm]{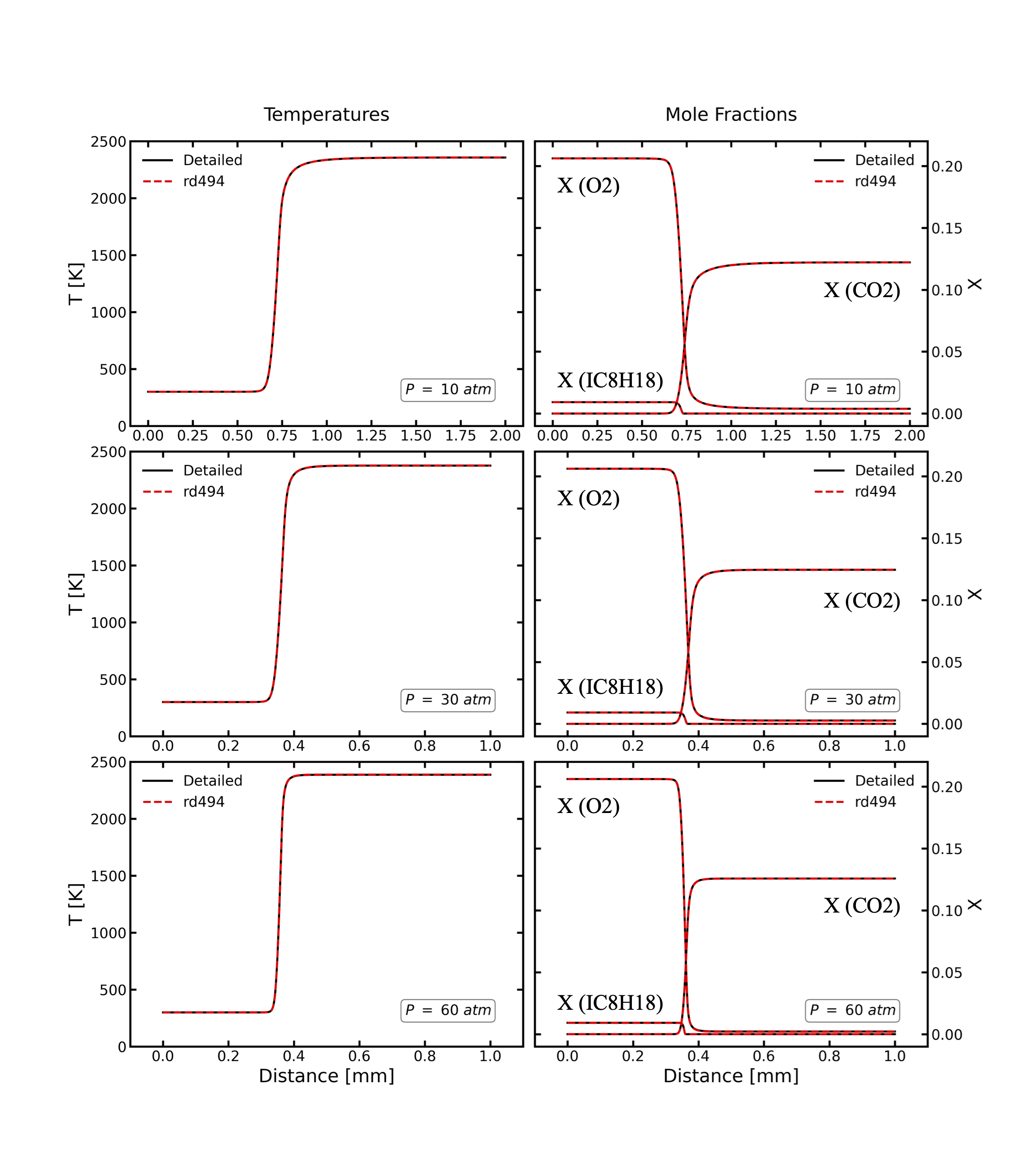}
 \caption{1D flame structure test of the $rd494$ model. Left: The distribution of flame temperatures. Right: Mole fractions of key species.}
\label{FIG:flamestructure}
\end{figure}

Simulations are also conducted to determine the laminar flame speed with $T_0 = 300$ $k$. Figure \ref{FIG:flamespeed} shows that all the relative errors are within 6\%. The $rd494$ model provides less than 3\% error in most cases, while $rd679$ yields less than 2\% error for all cases. The 1D flame structures are determined for several pressure values. In Fig.~\ref{FIG:flamestructure}, the distribution of temperatures and mole fractions of $O_2$, $CO_2$, and $IC_8H_{18}$ corroborate that the $rd494$ model perfectly replicates the 1D flame structure. 

\section{Conclusions}
\label{section:conclusion}
The implicit time-dependent basis CUR (implicit TDB-CUR) method is employed for low-rank approximation of linear sensitivities in the gasoline surrogate combustion based on the Lawrence Livermore National Laboratory (LLNL) kinetics model. It is shown that this method evaluates sensitivities with computational costs significantly lower than that required to solve the full-order model (FOM). The computed sensitivities are then analyzed for skeletal reduction of the LLNL surrogate fuel model. In doing so, the species are ranked based on their sensitivities to the reaction rates. The $rd679$ skeletal model as developed here shows the same accuracy as the $chem679$ model developed in Ref.~\cite{lu09tow}.
The $rd494$ with a lower number of species is also generated. The accuracy and efficiency of the model are assessed by comparing its predictions for the ignition delay, the flame speed, and 1-D flame structure. The $rd494$ model, which contains about 35\% of the total species suggested by the LLNL model, while yielding better than 90\% accuracy, is a reliable and affordable alternative to the detailed model.

This work demonstrates that the implicit TDB-CUR method accurately approximates the sensitivities and generates skeletal reaction models in an automated and efficient manner. The ability to perform such a reduction automatically for large and complex systems eliminates the need for user expertise in chemical kinetics networks, marking a significant advancement. Although the reduction scheme is applied to a specific gasoline surrogate mixture, it is broadly applicable to the reduction of other multi-component fuels. Moreover, it is well-suited for reducing other reaction networks, such as nuclear~\cite{nouri2019mod} and biochemical systems~\cite{chen2010cla}.

\section*{Acknowledgments}
This work is co-authored by an employee of Triad National Security, LLC which operates Los Alamos National Laboratory under Contract No.89233218CNA000001 with the U.S. Department of Energy/National Nuclear Security Administration. The work at Pitt is sponsored by the NSF under Grant CBET-2042918 and Grant CBET-2152803. Computational resources are provided by the Center for Research Computing (CRC) at Pitt.
\section*{Appendix}
\label{Appendix}
All of the reduced models can be accessed via this \href{https://github.com/Tom-Y-Liu/Chemical-Mechanisms.git}{\underline{link}}. 
\typeout{}
\bibliography{ROM_combustion, Gasoline, Methane_YL}

\begin{thebibliography}{10}
\expandafter\ifx\csname url\endcsname\relax
  \def\url#1{\texttt{#1}}\fi
\expandafter\ifx\csname urlprefix\endcsname\relax\def\urlprefix{URL }\fi
\expandafter\ifx\csname href\endcsname\relax
  \def\href#1#2{#2} \def\path#1{#1}\fi

\bibitem{zhen17ove}
X.~Zhen, Y.~Wang, D.~Liu, {A}n {O}verview of the {C}hemical {R}eaction {M}echanisms for {G}asoline {S}urrogate {F}uels, Appl. Therm. Eng. 124 (2017) 1257--1268.

\bibitem{sarathy18rec}
S.~M. Sarathy, A.~Farooq, G.~T. Kalghatgi, {R}ecent {P}rogress in {G}asoline {S}urrogate {F}uels, Prog. Energy Combust. Sci. 65 (2018) 67--108.

\bibitem{law10com}
C.~K. Law, {C}ombustion {P}hysics, {C}ambridge {U}niversity {P}ress, 2010.

\bibitem{de16rev}
L.~P. de~Oliveira, D.~Hudebine, D.~Guillaume, J.~J. Verstraete, {A} {R}eview of {K}inetic {M}odeling {M}ethodologies for {C}omplex {P}rocesses, Oil Gas Sci Technol–Revue d'IFP Energies nouvelles 71~(3) (2016) 45.

\bibitem{GM11}
D.~Goussis, U.~Maas, {M}odel {R}eduction for {C}ombustion {C}hemistry, in: T.~Echekki, E.~Mastorakos (Eds.), Turbulent {C}ombustion {M}odeling, Springer, 2011, pp. 193--220.

\bibitem{SCGJ10}
W.~Sun, Z.~Chen, X.~Gou, Y.~Ju, {A} {P}ath {F}lux {A}nalysis {M}ethod for the {R}eduction of {D}etailed {C}hemical {K}inetic {M}echanisms, Combust. Flame 157~(7) (2010) 1298--1307.

\bibitem{wang16imp}
W.~Wang, X.~Gou, {A}n {I}mproved {P}ath {F}lux {A}nalysis with {M}ulti {G}enerations {M}ethod for {M}echanism {R}eduction, Combust. Theory Model. 20~(2) (2016) 203--220.

\bibitem{lam89und}
S.~H. Lam, D.~A. Goussis, {U}nderstanding {C}omplex {C}hemical {K}inetics with {C}omputational {S}ingular {P}erturbation, Proc. Combust. Inst. 22~(1) (1989) 931--941.

\bibitem{lam93usi}
S.~H. Lam, {U}sing {CSP} to {U}nderstand {C}omplex {C}hemical {K}inetics, Combust. Sci. Technol. 89~(5-6) (1993) 375--404.

\bibitem{lam94csp}
S.~H. Lam, D.~A. Goussis, {T}he {CSP} {M}ethod for {S}implifying {K}inetics, Int. J. Chem. Kinet. 26~(4) (1994) 461--486.

\bibitem{lu01com}
T.~Lu, Y.~Ju, C.~K. Law, {C}omplex {CSP} for {C}hemistry {R}eduction and {A}nalysis, Combust. Flame 126~(1-2) (2001) 1445--1455.

\bibitem{LL05}
T.~Lu, C.~K. Law, {A} {D}irected {R}elation {G}raph {M}ethod for {M}echanism {R}eduction, Proc. Combust. Inst. 30~(1) (2005) 1333--1341.

\bibitem{ZLL07}
X.~Zheng, T.~Lu, C.~K. Law, {E}xperimental {C}ounterflow {I}gnition {T}emperatures and {R}eaction {M}echanisms of 1, 3-{B}utadiene, Proc. Combust. Inst. 31~(1) (2007) 367--375.

\bibitem{LL09}
T.~Lu, C.~Law, {T}oward {A}ccommodating {R}ealistic {F}uel {C}hemistry in {L}arge-{S}cale {C}omputations, Prog. Energy Combust. Sci. 35~(2) (2009) 192--215.

\bibitem{NSR10}
K.~Niemeyer, C.~Sung, M.~Raju, {S}keletal {M}echanism {G}eneration for {S}urrogate {F}uels using {D}irected {R}elation {G}raph with {E}rror {P}ropagation and {S}ensitivity {A}nalysis, Combust. Flame 157~(9) (2010) 1760--1770.

\bibitem{Shuang16}
S.~Li, B.~Yang, F.~Qi, {A}ccelerate {G}lobal {S}ensitivity {A}nalysis using {A}rtificial {N}eural {N}etwork {A}lgorithm: {C}ase {S}tudies for {C}ombustion {K}inetic {M}odel, Combust. Flame 168 (2016) 53--64.

\bibitem{Turanyi90b}
T.~Tur{\'a}nyi, {S}ensitivity {A}nalysis of {C}omplex {K}inetic {S}ystems. {T}ools and {A}pplications, J. Math. Chem. 5~(3) (1990) 203--248.

\bibitem{Sobol90}
I.~Sobol', {O}n {S}ensitivity {E}stimation for {N}onlinear {M}athematical {M}odels, Matematicheskoe Modelirovanie 2~(1) (1990) 112--118.

\bibitem{HS96}
T.~Homma, A.~Saltelli, {I}mportance {M}easures in {G}lobal {S}ensitivity {A}nalysis of {N}onlinear {M}odels, Reliab. Eng. Syst. Saf. 52~(1) (1996) 1--17.

\bibitem{RA99}
H.~Rabitz, {\"O}.~Ali{\c{s}}, {G}eneral {F}oundations of {H}igh-{D}imensional {M}odel {R}epresentations, J. Math. Chem. 25~(2-3) (1999) 197--233.

\bibitem{Sobol01}
I.~Sobol', {G}lobal {S}ensitivity {I}ndices for {N}onlinear {M}athematical {M}odels and {T}heir {M}onte {C}arlo {E}stimates, Math. Comput. Simul. 55~(1-3) (2001) 271--280.

\bibitem{LSR02}
G.~Li, S.~Wang, H.~Rabitz, {P}ractical {A}pproaches to {C}onstruct {RS-HDMR} {C}omponent {F}unctions, J. Phys. Chem. A 106~(37) (2002) 8721--8733.

\bibitem{Saltelli04}
A.~Saltelli, S.~Tarantola, F.~Campolongo, M.~Ratto, {S}ensitivity {A}nalysis in {P}ractice: {A} {G}uide to {A}ssessing {S}cientific {M}odels, John Wiley \& Sons, 2004.

\bibitem{SYWL09}
D.~A. Sheen, X.~You, H.~Wang, T.~L{\o}v{\aa}s, {S}pectral {U}ncertainty {Q}uantification, {P}ropagation and {O}ptimization of a {D}etailed {K}inetic {M}odel for {E}thylene {C}ombustion, Proc. Combust. Inst. 32~(1) (2009) 535--542.

\bibitem{ESC12}
G.~Esposito, B.~Sarnacki, H.~Chelliah, {U}ncertainty {P}ropagation of {C}hemical {K}inetics {P}arameters and {B}inary {D}iffusion {C}oefficients in {P}redicting {E}xtinction {L}imits of {H}ydrogen/{O}xygen/{N}itrogen {N}on-premixed {F}lames, Combust. Theory Model. 16~(6) (2012) 1029--1052.

\bibitem{lakrisenko24ben}
P.~Lakrisenko, D.~Pathirana, D.~Weindl, J.~Hasenauer, {B}enchmarking {M}ethods for {C}omputing {L}ocal {S}ensitivities in {O}rdinary {D}ifferential {E}quation {M}odels at {D}ynamic and {S}teady {S}tates, PLOS ONE 19~(10) (2024) e0312148.

\bibitem{lakrisenko23eff}
P.~Lakrisenko, P.~Stapor, S.~Grein, {\L}.~Paszkowski, D.~Pathirana, F.~Fr{\"o}hlich, G.~T. Lines, D.~Weindl, J.~Hasenauer, {E}fficient {C}omputation of {A}djoint {S}ensitivities at {S}teady-state in {ODE} {M}odels of {B}iochemical {R}eaction {N}etworks, PLOS Comput. Biol. 19~(1) (2023) e1010783.

\bibitem{BLK97}
N.~Brown, G.~Li, M.~Koszykowski, {M}echanism {R}eduction via {P}rincipal {C}omponent {A}nalysis, Int. J. Chem. Kinet. 29~(6) (1997) 393--414.

\bibitem{EC11}
G.~Esposito, H.~Chelliah, {S}keletal {R}eaction {M}odels based on {P}rincipal {C}omponent {A}nalysis: {A}pplication to {E}thylene--{A}ir {I}gnition, {P}ropagation, and {E}xtinction {P}henomena, Combust. Flame 158~(3) (2011) 477--489.

\bibitem{PSDTS11}
A.~Parente, J.~Sutherland, B.~Dally, L.~Tognotti, P.~Smith, {I}nvestigation of the {MILD} {C}ombustion {R}egime via {P}rincipal {C}omponent {A}nalysis, Proc. Combust. Inst. 33~(2) (2011) 3333--3341.

\bibitem{PS13}
A.~Parente, J.~Sutherland, {P}rincipal {C}omponent {A}nalysis of {T}urbulent {C}ombustion {D}ata: {D}ata {P}re-processing and {M}anifold {S}ensitivity, Combust. Flame 160~(2) (2013) 340--350.

\bibitem{ME14}
H.~Mirgolbabaei, T.~Echekki, {N}onlinear {R}eduction of {C}ombustion {C}omposition {S}pace with {K}ernel {P}rincipal {C}omponent {A}nalysis, Combust. Flame 161~(1) (2014) 118--126.

\bibitem{CIGP16}
A.~Coussement, B.~Isaac, O.~Gicquel, A.~Parente, {A}ssessment of {D}ifferent {C}hemistry {R}eduction {M}ethods based on {P}rincipal {C}omponent {A}nalysis: {C}omparison of the {MG-PCA} and {S}core-{PCA} {A}pproaches, Combust. Flame 168 (2016) 83--97.

\bibitem{MICSP18}
M.~Malik, B.~Isaac, A.~Coussement, P.~Smith, A.~Parente, {P}rincipal {C}omponent {A}nalysis {C}oupled with {N}onlinear {R}egression for {C}hemistry {R}eduction, Combust. Flame 187 (2018) 30--41.

\bibitem{SFCFR16}
A.~Stagni, A.~Frassoldati, A.~Cuoci, T.~Faravelli, E.~Ranzi, {S}keletal {M}echanism {R}eduction {T}hrough {S}pecies-{T}argeted {S}ensitivity {A}nalysis, Combust. Flame 163 (2016) 382--393.

\bibitem{NBGCL22}
A.~Nouri, H.~Babaee, P.~Givi, H.~Chelliah, D.~Livescu, {S}keletal {M}odel {R}eduction with {F}orced {O}ptimally {T}ime {D}ependent {M}odes, Combust. Flame 235 (2022) 111684.

\bibitem{LBGCLN24}
Y.~Liu, H.~Babaee, P.~Givi, H.~Chelliah, D.~Livescu, A.~Nouri, {S}keletal {R}eaction {M}odels for {M}ethane {C}ombustion, Fuel 357 (2024) 129581.

\bibitem{nouri24ske}
A.~Nouri, Y.~Liu, P.~Givi, H.~Babaee, D.~Livescu, {S}keletal {K}inetics {R}eduction for {A}strophysical {R}eaction {N}etworks, Astrophys. J. Suppl. S. 272~(2) (2024) 34.

\bibitem{DPNFB23}
M.~Donello, G.~Palkar, M.~Naderi, D.~Del Rey~Fern{\'a}ndez, H.~Babaee, {O}blique {P}rojection for {S}calable {R}ank-adaptive {R}educed-order {M}odelling of {N}onlinear {S}tochastic {P}artial {D}ifferential {E}quations with {T}ime-dependent {B}ases, Proc. R. Soc. A 479~(2278) (2023) 20230320.

\bibitem{naderi25cro}
M.~H. Naderi, S.~Akhavan, H.~Babaee, {A} {C}ross {A}lgorithm for {I}mplicit {T}ime {I}ntegration of {R}andom {P}artial {D}ifferential {E}quations on {L}ow-rank {M}atrix {M}anifolds, Proc. R. Soc. A 481~(2309) (2025) 20240658.

\bibitem{goreinov1997the}
S.~A. Goreinov, E.~E. Tyrtyshnikov, N.~L. Zamarashkin, {A} {T}heory of {P}seudoskeleton {A}pproximations, Linear Algebra Its Appl. 261~(1-3) (1997) 1--21.

\bibitem{mahoney2009cur}
M.~W. Mahoney, P.~Drineas, {CUR} {M}atrix {D}ecompositions for {I}mproved {D}ata {A}nalysis, Proc. Natl. Acad. Sci. U.S.A. 106~(3) (2009) 697--702.

\bibitem{mehl11det}
M.~Mehl, W.~J. Pitz, C.~K. Westbrook, H.~J. Curran, {K}inetic {M}odeling of {G}asoline {S}urrogate {C}omponents and {M}ixtures {U}nder {E}ngine {C}onditions, Proc. Combust. Inst. 33~(1) (2011) 193--200.

\bibitem{kuk2012exp}
G.~Kukkadapu, K.~Kumar, C.-J. Sung, M.~Mehl, W.~J. Pitz, {E}xperimental and {S}urrogate {M}odeling {S}tudy of {G}asoline {I}gnition in {A} {R}apid {C}ompression {M}achine, Combust. Flame 159~(10) (2012) 3066--3078.

\bibitem{kuk2013aut}
G.~Kukkadapu, K.~Kumar, C.-J. Sung, M.~Mehl, W.~J. Pitz, {A}utoignition of {G}asoline and {I}ts {S}urrogates in {A} {R}apid {C}ompression {M}achine, Proc. Combust. Inst. 34~(1) (2013) 345--352.

\bibitem{mehl2011red}
M.~Mehl, J.-Y. Chen, W.~J. Pitz, S.~M. Sarathy, C.~K. Westbrook, {A}n {A}pproach for {F}ormulating {S}urrogates for {G}asoline {W}ith {A}pplication {T}oward a {R}educed {S}urrogate {M}echanism for {CFD} {E}ngine {M}odeling, Energy \& Fuels 25~(11) (2011) 5215--5223.

\bibitem{lu09tow}
T.~Lu, C.~K. Law, {T}oward {A}ccommodating {R}ealistic {F}uel {C}hemistry in {L}arge-scale {C}omputations, Prog. Energy Combust. Sci. 35~(2) (2009) 192--215.

\bibitem{niemeyer15red}
K.~E. Niemeyer, C.-J. Sung, {R}educed {C}hemistry for {A} {G}asoline {S}urrogate {V}alid at {E}ngine-relevant {C}onditions, Energy \& Fuels 29~(2) (2015) 1172--1185.

\bibitem{curtiss52int}
C.~F. Curtiss, J.~O. Hirschfelder, {I}ntegration of {S}tiff {E}quations, Proc. Natl. Acad. Sci. U.S.A. 38~(3) (1952) 235--243.

\bibitem{gear1967num}
C.~W. Gear, {T}he {N}umerical {I}ntegration of {O}rdinary {D}ifferential {E}quations, Math. Comput. 21~(98) (1967) 146--156.

\bibitem{suli2003int}
E.~S{\"u}li, D.~F. Mayers, {A}n {I}ntroduction to {N}umerical {A}nalysis, Cambridge university press, 2003.

\bibitem{ZZT03}
I.~Zsely, J.~Zador, T.~Turanyi, {S}imilarity of {S}ensitivity {F}unctions of {R}eaction {K}inetic {M}odels, J. Phys. Chem. A 107~(13) (2003) 2216--2238.

\bibitem{PDG20}
B.~Peherstorfer, Z.~Drmac, S.~Gugercin, {S}tability of {D}iscrete {E}mpirical {I}nterpolation and {G}appy {P}roper {O}rthogonal {D}ecomposition with {R}andomized and {D}eterministic {S}ampling {P}oints, SIAM J Sci Comput 42~(5) (2020) A2837--A2864.

\bibitem{ghiasi2019gas}
G.~Ghiasi, I.~Ahmed, N.~Swaminathan, {G}asoline {F}lame {B}ehavior at {E}levated {T}emperature and {P}ressure, Fuel 238 (2019) 248--256.

\bibitem{law2012ntc}
C.~K. Law, P.~Zhao, {NTC}-affected {I}gnition in {N}onpremixed {C}ounterflow, Combust. Flame 159~(3) (2012) 1044--1054.

\bibitem{nouri2019mod}
A.~G. Nouri, P.~Givi, D.~Livescu, {M}odeling {A}nd {S}imulation of {T}urbulent {N}uclear {F}lames in {T}ype {I}a {S}upernovae, Prog. Aerosp. Sci. 108 (2019) 156--179.

\bibitem{chen2010cla}
W.~W. Chen, M.~Niepel, P.~K. Sorger, {C}lassic and {C}ontemporary {A}pproaches to {M}odeling {B}iochemical {R}eactions, Genes Dev. 24~(17) (2010) 1861--1875.

\end{thebibliography}

\end{document}